\documentclass[aps,pra,twocolumn,superscriptaddress]{revtex4}
\usepackage{graphicx,float}
\usepackage{epsfig}
\usepackage{dcolumn}
\usepackage{ulem}
\usepackage{bm}
\usepackage{amsmath}
\usepackage{amssymb}
\usepackage{hyperref}
\hypersetup{colorlinks=true,linkcolor=blue,citecolor=blue,urlcolor=black}

\usepackage{csquotes}
\usepackage{mathtools}
\usepackage{physics}
\usepackage{soul}

\begin{document}
%%%%%%%%%%%%%%%%%%%%%%%%%%%%%%%%%%%%%%%%%%%%%%%%%%%%%%%%%%%%%%%%%

\title{Multi-core fiber integrated multi-port beamsplitters for quantum information processing}
%\title{Multi-port beamsplitters based on multi-core optical fibers for high-dimensional quantum information}

\author{J.~Cari\~{n}e}
\affiliation{Departamento de F\'{\i}sica, Universidad de Concepci\'on, 160-C Concepci\'on, Chile}
\affiliation{Millennium Institute for Research in Optics, Universidad de Concepci\'on, 160-C Concepci\'on, Chile}
\author{G.~Ca{\~n}as}
\affiliation{Departamento de F\'{\i}sica, Universidad del Bio-Bio, Avenida Collao 1202, Concepci\'on, Chile}
\author{P.~Skrzypczyk}
\affiliation{H. H. Wills Physics Laboratory, University of Bristol, Tyndall Avenue, Bristol, BS8 1TL, United Kingdom}
\author{I.~\v{S}upi\'{c}}
\affiliation{ICFO-Institut de Ciencies Fotoniques, The Barcelona Institute of Science and Technology, 08860 Castelldefels, Barcelona, Spain}
\author{N.~Guerrero}
\affiliation{Departamento de F\'{\i}sica, Universidad de Concepci\'on, 160-C Concepci\'on, Chile}
\affiliation{Millennium Institute for Research in Optics, Universidad de Concepci\'on, 160-C Concepci\'on, Chile}
\author{T.~Garcia}
\affiliation{Departamento de F\'{\i}sica, Universidad de Concepci\'on, 160-C Concepci\'on, Chile}
\affiliation{Millennium Institute for Research in Optics, Universidad de Concepci\'on, 160-C Concepci\'on, Chile}
\author{L.~Pereira}
\affiliation{Departamento de F\'{\i}sica, Universidad de Concepci\'on, 160-C Concepci\'on, Chile}
\affiliation{Millennium Institute for Research in Optics, Universidad de Concepci\'on, 160-C Concepci\'on, Chile}
\author{M.~A.~S.-Prosser}
\affiliation{Departamento de Ciencias F\'{\i}sicas, Universidad de la Frontera, Temuco, Chile}
\author{G.~B.~Xavier}
\affiliation{Institutionen f\"{o}r Systemteknik, Link\"{o}pings Universitet, 581 83 Link\"{o}ping, Sweden}
\author{A.~Delgado}
\affiliation{Departamento de F\'{\i}sica, Universidad de Concepci\'on, 160-C Concepci\'on, Chile}
\affiliation{Millennium Institute for Research in Optics, Universidad de Concepci\'on, 160-C Concepci\'on, Chile}
\author{S.~P.~Walborn}
\affiliation{Departamento de F\'{\i}sica, Universidad de Concepci\'on, 160-C Concepci\'on, Chile}
\affiliation{Millennium Institute for Research in Optics, Universidad de Concepci\'on, 160-C Concepci\'on, Chile}
\affiliation{Instituto de F\'{i}sica, Universidade Federal do Rio de Janeiro, Caixa Postal 68528, Rio de Janeiro, Rio de Janeiro 21941-972, Brazil}
\author{D.~Cavalcanti}
\affiliation{ICFO-Institut de Ciencies Fotoniques, The Barcelona Institute of Science and Technology, 08860 Castelldefels, Barcelona, Spain}
\author{G.~Lima}
\affiliation{Departamento de F\'{\i}sica, Universidad de Concepci\'on, 160-C Concepci\'on, Chile}
\affiliation{Millennium Institute for Research in Optics, Universidad de Concepci\'on, 160-C Concepci\'on, Chile}

%%%%%%%%%%%%%%%%%%%%%%%%%%%%%%%%%%%%%%%%%%%%%%%%%%%%%%%%%%%%%%%%%

\begin{abstract}
Multi-port beamsplitters are cornerstone devices for high-dimensional quantum information tasks, which can outperform the two-dimensional ones. Nonetheless, the fabrication of such devices has been proven to be challenging with progress only recently achieved with the advent of integrated photonics. Here, we report on the production of high-quality $N \times N$ (with $N=4,7$) multi-port beamsplitters based on a new scheme for manipulating multi-core optical fibers. By exploring their compatibility with optical fiber components, we create 4-dimensional quantum systems and implement the measurement-device-independent random number generation task with a programmable 4-arm interferometer operating at a 2 MHz repetition rate. Thanks to the high visibilities observed, we surpass the 1-bit limit of binary protocols and attain 1.23 bits of certified private randomness per experimental round. Our result demonstrates that fast switching, low-loss and high optical quality for high-dimensional quantum information can be simultaneously achieved with multi-core fiber technology.
\end{abstract}

%%%%%%%%%%%%%%%%%%%%%%%%%%%%%%%%%%%%%%%%%%%%%%%%%%%%%%%%%%%%%%%%%%%

%\setboolean{displaycopyright}{true}

%\begin{document}

\maketitle

%%%%%%%%%%%%%%%%%%%%%%%%%%%%%%%%%%%%%%%%%%%%%%%%%%%%%%%%%%%%%%%%%%%

\section{Introduction}
\label{sec:intro}

Space-division multiplexing (SDM) is currently the main technology considered to overcome the actual capacity limitation of optical telecommunication networks \cite{Richardson_natphoton_2013}. Basically, it consists of specially designed fibers that can support distinct optical spatial modes in order to increase the multiplexing capabilities. The optical fibers employed in SDM can be divided into two main groups: multi-core fibers (MCF) \cite{Iano_1979,Saitoh_2016} and few-mode fibers (FMF) \cite{Sillard_2014,Bozinovic_2013,Brunet_2015,Gregg_2015,Wang}. In the former, several single-mode cores are physically contained within the same common cladding, with each core being used independently. A FMF on the other hand consists of a single core that supports several optical modes, each of them capable of transmitting data independently.

Arguably, the development of a major part of experimental quantum information (QI) relies on the fact that it is heavily based on the same hardware employed by classical optical communication \cite{Gisin_2002,Lo_2014,Diamanti_2016,Xu_2019,Pirandola_2019}. Therefore, it is natural to expect that future development will take place using SDM hardware \cite{GuixReview_2019}. Indeed, in the past couple of years the first quantum communication experiments based on MCFs have appeared. The first one used a MCF as a direct multiplexing device: with one core acting as the quantum channel, while other cores contained classical data \cite{Toshiba2016}. See also Refs. \cite{Lin_2018,Hugues-Salas_2019,Cai_2019,Eriksson_2019}. Later, the fact that all cores are placed in a common cladding translates to a long multi-path conduit with intrinsic phase stability, was explored for demonstrating the feasibility of high-dimensional (HD) quantum key distribution over MCFs \cite{Canas_2017,Ding_2017}. The relative phase difference between multiple cores of MCF fibers has been shown to be more stable than that of multiple single-mode fibers by at least two orders of magnitude over a 2 km fiber link \cite{Lio_2020}. The benefit of MCFs for QI has been further reinforced by showing that they can support propagation of entangled photons \cite{Lee_2017,Lee_2019}. Similar research has begun for FMFs \cite{Cui_2017,Sit_2018,Cao_2018,Cozzolino_2019,Liu_2019,Valencia_2019}. HD entanglement is advantageous in this regard, as it can be more resistant to noise \cite{Ecker_2019}.

Additionally, SDM technology has been exploited for building MCF based optical fiber sensors, whose remote interrogation capabilities makes them attractive for industrial applications \cite{Amezcua,PingShum,Libo,Xuewen,Ming}. MCF optical sensors have been used for high-temperature sensing up to 1000 $^\circ$C with a typical temperature sensitivity as high as 170 pm/$^\circ$C \cite{Xuewen}. The advantage of using MCFs is that they allow for the fabrication of multi-arm Mach-Zehnder (MZ) interferometers that have higher sensitivity for phase changes since the slopes of the resulting interference peaks are steeper. There has been a large variety of MCF optical sensors but most of them rely on inefficient techniques to launch light into the multi-arm MZ, resulting in prohibitive losses for quantum information processing. Of particular interest is the work of L. Gan et. al. \cite{Ming}, where the authors develop new tapering techniques to build the multi-arm MZ directly into a specially designed MCF.

Inspired by such progress in optical sensing, we report on the production of high-quality $N \times N$ (with $N=4,7$) multi-port beamsplitters (MBS) built-in commercially available multi-core fibers and their usage for building fast, low-loss and programable multi-arm MZ interferometers suitable for QI. In the field of quantum computing, optical interferometers have attracted much attention. Since the seminal work of Knill, Laflamme and Milburn \cite{klm01}, it has been known that one possible road to universal quantum computing is through an architecture composed of single-photon sources, detectors, and linear-optic multi-arm interferometers. Such interferometers work as quantum circuits that are especially relevant for the efficient processing of HD photonic quantum systems (qudits), whose generation has now been harnessed \cite{Neves_2005,Boyd_2005,Kwiat2005,Groblacher_2006,QKD_Steve,Rossi_2009,Gao2010,Dada2011,QKD16,Spadua_16,Malik2016,Bristol2018,Aguilar_2018,Armin_2018}. Nonetheless. the development of MBS devices has been proven to be challenging \cite{Zeilinger95}. Recent progress has been made with the advent of integrated photonics \cite{Ding_2017,Politi_2008,Obrien2015,FabioTritter,Bristol2018}, where multi-arm interferometers are built resorting to a mesh of conventional $2 \times 2$ beamsplitters \cite{Reck,Zukowski}. In this case, the circuits can present balanced and unbalanced losses, and depending on the circuit size, the fidelity of the operations can be compromised \cite{Walmsley,FabioBenchmarking}. By taking a new approach based on MCFs for building multi-arm interferometers we present both: (i) a new technology that has technical advantages and is fully compatible with previous efforts in integrated photonics \cite{Ding_2017}, and that at the same time (ii) can be independently used for the high-quality processing of quantum information. It allows one to exploit the stability and compactness of MCF fibers, and their compatibility with trends in telecommunication technology, to build new robust schemes for optical sensing, communication and information processing. Note that $N$-arm interferometers can also be built with $2 \times 2$ in-fiber beamsplitters, but the scaling of the quantum circuit favours the use of MBSs. While two $N\times N$ MBSs suffice for a large class of transformations, the number of $2\times 2$ 50:50 beamsplitters is $N(N-1)$ \cite{Reck,Saygin_2020,PUMA}.

To demonstrate the viability of our approach for HD-QI, we consider the task of random number generation (RNG), which finds several applications in cryptography, gambling, and numerical simulations. In the classical domain, randomness is associated to our ignorance about the parameters describing a process. This perspective is not enough for cryptographic protocols, where we would like to certify that certain data is random for an eavesdropper that could have more knowledge or computational power than the user \cite{AcinMasanes_2017}. This problem was solved by fully device-independent (DI) RNG protocols \cite{AcinBELLNATURE}, which makes no assumptions on the source or measurements being used in the protocol \cite{NL_review}. However, this approach is quite demanding and typically results in very low random bit rates (see e.g. \cite{Pan2018}). A solution is to consider semi-device independent scenarios \cite{Marcin1,Marcin2,Brunner1,Brunner2,Himbeeck_2017,Rusca_2019}, where partial knowledge on the implementation is assumed. In our scheme, we assume that we control the source of quantum states but do not assume anything about the measurements we perform, a situation called measurement-device-independent (MDI) RNG \cite{Daniel2017}. Our implementation resorted to a MCF-based 4-arm interferometer operating at 2 MHz repetition rate, which generates and measures path encoded 4-dimensional qudit states with fidelities higher than 99.4$\%$. Moreover, we employ theoretical techniques that allow us to handle the issues with finite statistics, and use semi-definite programming to estimate the randomness in this MDI setting. This allowed us to attain a generation rate of 1.23 random bits per experimental round, which surpasses the 1-bit limit of binary RNG protocols. Thus, proving the usefulness of exploiting qudit states for RNG.

{Lastly}, we note that the average insertion loss for the fabricated 4$\times$4 (7$\times$7) MBSs is only $4.3\%$ ($9.0\%$), which allows for a qudit transmission of 42$\%$ through the programmable circuit and a corresponding overall detection efficiency that can reach at least 35$\%$ with commercially available superconducting single-photon detectors (efficiency $> 85\%$). This result, together with the interferometer's fast switching and high optical quality, yields potential advantages of this technology for quantum communication \cite{Cerf,Kaszlikowski_2000}, sensing \cite{Pirandola} and computation \cite{Obrien10,Araujo}.

\section{Fabrication, modelling and validation of the multi-port beamsplitters}
\label{sec:validation}

As mentioned before, the cladding of an MCF is composed of several cores, which can be exploited for the propagation of path qudit states defined as the coherent superposition {$|\Psi\rangle = \frac{1}{\sqrt{k}}\sum_{0}^{k}e^{i\phi_k}|k\rangle$ \cite{Canas_2017}, where $|k\rangle$ denotes the state of the photon transmitted by the $k$th core mode, and $\phi_k$ is the relative phase acquired during propagation over the $k$th core (See Fig.~\ref{Fig1}a). The high-quality $4 \times 4$ multi-port beamsplitters are constructed directly in a 4-core optical fiber through a tapering technique recently introduced in \cite{Ming}. In that work, the authors were interested in building multi-arm MZ interferometers for multi-parameter estimation. Their idea was to use a heterogeneous multicore fiber. This fiber is used to minimise inter-core coupling, as it has lower refractive-index “trenches” around the cores. In such fibers, there are at least two orthogonal modes propagating over one core of the fiber, which normally never interfere. Nonetheless, by tapering this fiber, they created an overlap between such modes due to strong evanescence effects in the tapered zone. From the interference observed, parameter estimation was possible. The authors then used each core interference for estimating different parameters of a sample. The fiber worked as an instrument composed of several 2-path MZ interferometers. In their tapered region, the inter-coupling between different cores was severely reduced by such trenches.

\begin{figure}[t]
\centering
\includegraphics[width=0.65\linewidth]{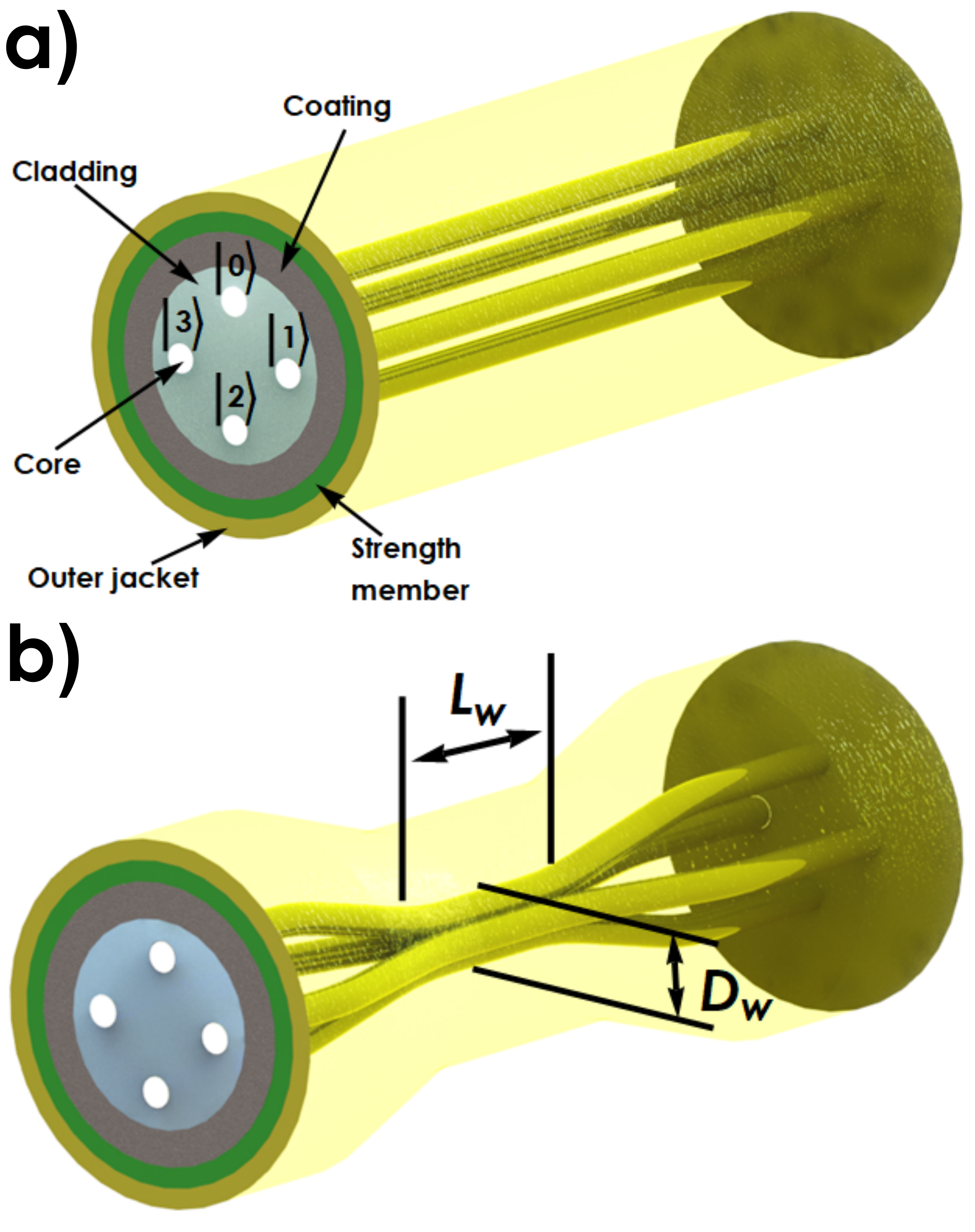}
\caption{Schematics of a MCF and of the fabricated MBSs. a) MCF before tapering and the qudit encoding strategy. b) The fiber is then heated along a length $L$ and pulled symmetrically from both ends, stretching and thinning the fiber. The final device is the MBS and has a length $L_W$ with diameter $D_W$. \label{Fig1}}
\end{figure}

Here, we show that by employing the same technique but with homogeneous MCFs, i.e., fibers where the $N$ cores are not bounded by refractive index trenches, one can build high-quality $N \times N$ multiport beamsplitters. The tapering is performed by locally heating a small transverse region of the fiber with length $L$, while applying a controlled longitudinal stretching tension. Since the fiber is mechanically in a partial soft state, it will become thinner with a final diameter $D_w$ at the center of the region where the heat is applied. The cores will consequently be brought together, and due to evanescent coupling, light will leak from one core to the others, similar to what is obtained in a standard fiber-optical bi-directional coupler (See Fig.~\ref{Fig1}b). The splitting ratio can be balanced by monitoring the transmission of a 1550nm laser beam sent through the 4-core fiber while tapering it. Finally, since the MBS is directly constructed on a MCF, it is compatible for connection with other MCFs by direct contact.

We test the fabricated 4-core MBSs by first illuminating one of the cores of a MCF. This fiber is connected to the MBS under test, and at the output, the light is split across the other cores. Figure \ref{Fig2}a shows the image of the output facet of one MBS on an infrared CCD camera, clearly showing the 4-core pattern, as well as the cores fully illuminated. We then measure the output power per core individually with p-i-n photodiodes. Figure \ref{Fig2}b shows the normalized intensity at each core following the MBS and its evolution over time. The power at each core is very stable and the observed average split ratio is ($0.248 \pm 0.01$). The average insertion loss of the 4$\times$4 MBSs is $(4.3 \pm 0.06)\%$.

\begin{figure}[t]
\centering
\includegraphics[width=0.9\linewidth]{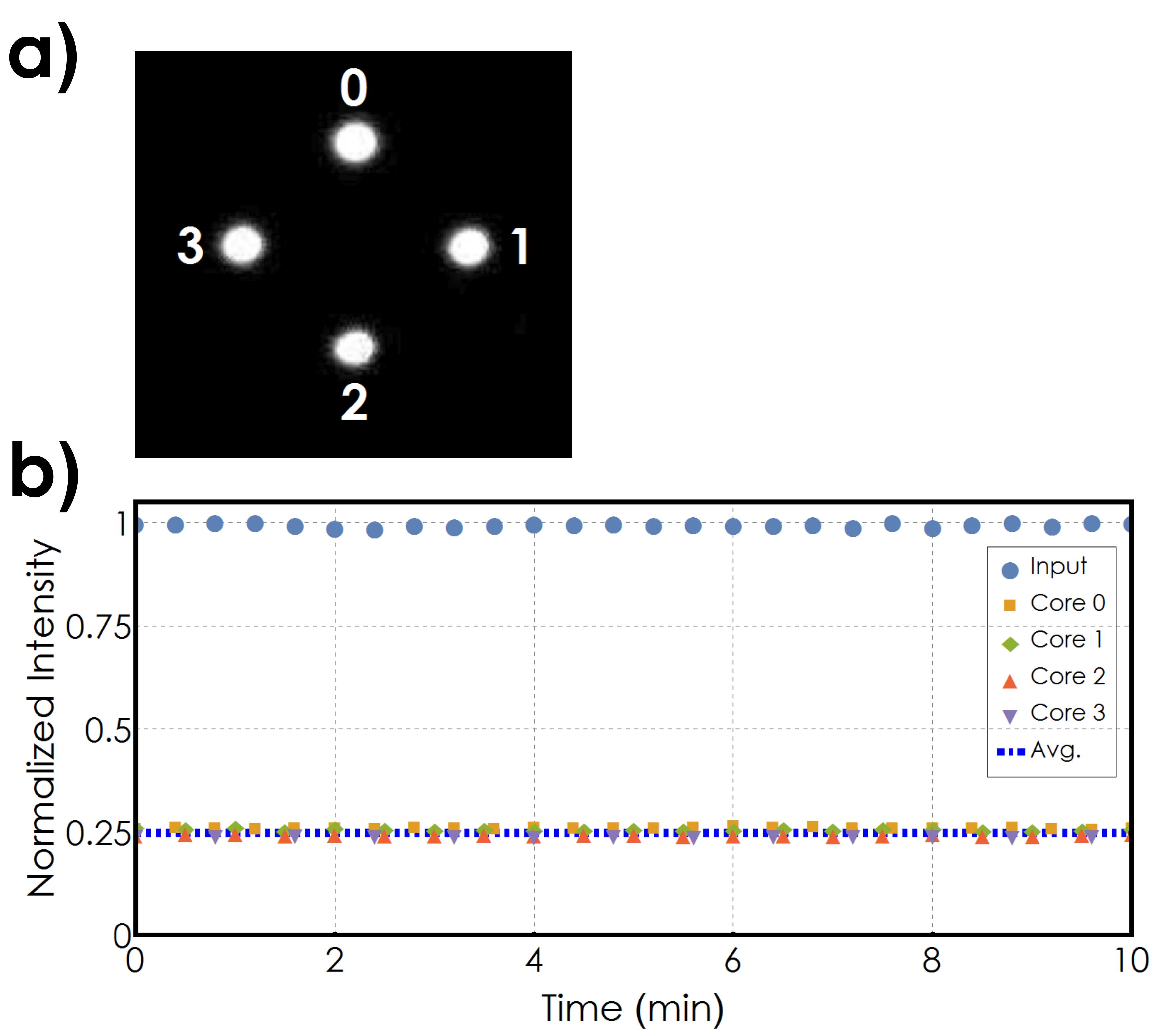}
\caption{Multi-port beamsplitter performance. a) Image of the facet of the output of a MCF 4$\times$4 MBS as seen by an infrared CCD camera. b) Output normalized optical power of each core of the MBS as a function of time. \label{Fig2}}
\end{figure}

In general, symmetric $4\times4$ multi-port BSs are parameterized in terms of the unitary operation given by \cite{Zeilinger95,Reck,Zukowski}
\vspace{-0.2cm}
\begin{equation}V=\frac{1}{2}
\begin{bmatrix}
 1 & 1 & 1 & 1 \\
 1 & e^{i\phi} & -1 & -e^{i\phi} \\
 1 & -1 & 1 & -1 \\
 1 & -e^{i\phi} &  -1 & e^{i\phi} \label{Matrix_Theo}
\end{bmatrix}.\end{equation} Since the cores are equally distant to the center of the 4-core MCF, in the tapered zone they will have the same length $L_w$. So, it is expected that the MCF MBSs should be described by $V$ when $\phi=0$. We confirm this by experimentally measuring the unitary implemented by a 4$\times$4 MCF MBS, resorting to the quantum process tomography technique introduced in \cite{QPTMBS}. Any unitary device is described by $U=\sum_{jk}u_{jk}e^{i\phi_{jk}}\dyad{j}{k}$. The parameters $ u_{jk}$ for our MBS are obtained from the split ratios recorded in the procedure described above. The relative phases are measured by sending states of the form $\ket{\phi_j}=\frac{1}{\sqrt{2}}\left( \ket{1}+e^{i\varphi}\ket{j}\right)$ through the MBS. At the MBS output ports, the probabilities of recording the photon are given by  $p(k|j)= \frac{1}{2}\left[ u_{k1}^2+u_{kj}^2+2u_{k1}u_{kj}\cos(\varphi+\phi_{kj}-\phi_{k1})\right]$. Hence, by recording these probabilities with respect to $\varphi$, {we} acquire the relative phases $\phi_{kj}-\phi_{k1}$. Using the scheme of Fig.~\ref{Fig3} explained below, we obtain the experimental matrix $\tilde U$. Nonetheless, due to inherent experimental errors, this matrix is never unitary. In order to obtain the unitary matrix describing {the} 4$\times$4 MCF MBS, one can optimize a cost function of the experimental data. For this purpose, the fidelity between two matrices, given by $F(A,B) = \frac{1}{N^2}\left| \Tr(A^\dagger B)\right|^2$ \cite{cite_fid1,cite_fid2}, is typically used as a figure of merit. Then, the final MBS matrix is given by the optimization problem: $\hat U = \arg \min_{V}\left[1-F(\tilde U,V)\right]$. Following this procedure, we determine that our MCF MBS matrix representation is
\vspace{-0.2cm}
\begin{equation}\hat U\!=\!
\begin{bmatrix}
0.499 &0,501 &0,499 &0,499 \\
0,501 &0,491\!+\!0,08i &-0,496\!-\!0,06i &-0,498\!-\!0,01i \\
0,499 &-0,495\!-\!0,06i &0,498\!+\!0,03i &-0,499\!+\!0,03i \\
0,499 &-0,499\!-\!0,01i &-0,499\!+\!0,03i &0,499\!-\!0,01i
\label{Matrix_Exp}
\end{bmatrix}, \end{equation} which has a fidelity with the model of Eq.~(\ref{Matrix_Theo}) given by $F(\hat U,V_{\phi=0})= 0.995\pm0.003$, confirming the high quality of the 4$\times$4 MCF MBSs. Last, we note that our technique can be extended to MCFs of more cores for creating multi-port BS with more input/output ports. We present the characterization of a 7$\times$7 MCF MBS in the supplementary material. The average insertion loss of that 7$\times$7 MBSs is $(9.0 \pm 0.04)\%$.

%************************************* FIG 3 ****************************************
\begin{figure*}[t]
\centering
\includegraphics[width=\linewidth]{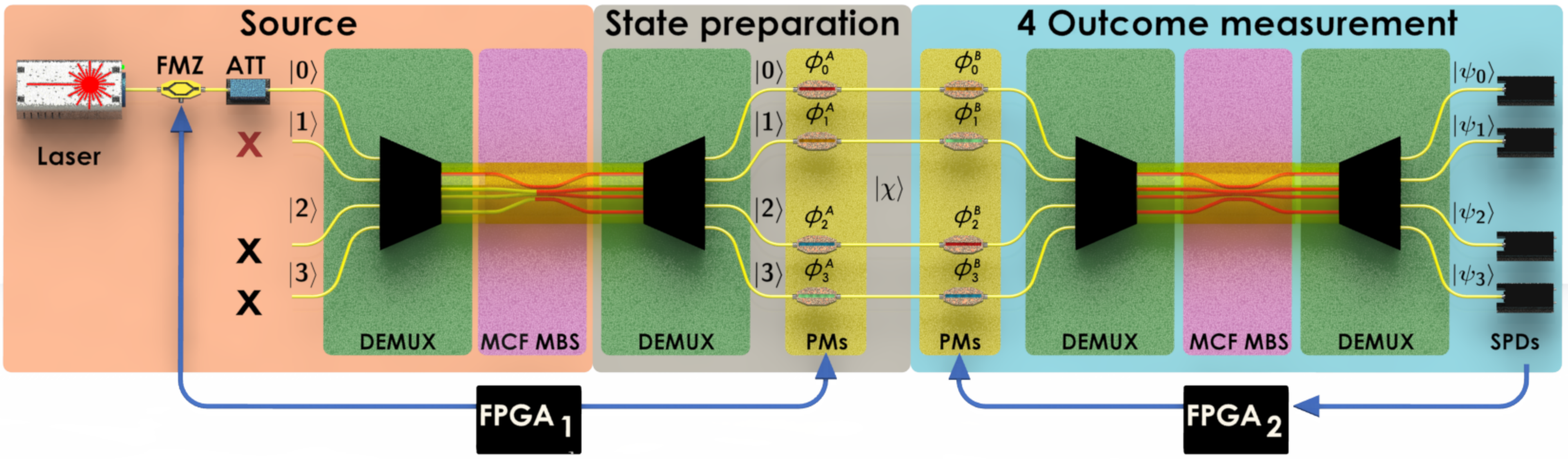}
\caption{Schematics of the experimental setup implementing the programmable quantum circuit for HD quantum information processing. Please see the main text for details.} \label{Fig3}
\end{figure*}
%*************************************************************************************

\section{Multi-arm interferometers based on multi-core fibers}

A programmable quantum circuit allows one to prepare different quantum states and measure them with different bases in a controllable way. Now, we show (i) how the MCF MBSs can be used to build a multi-arm MZ interferometer, and (ii) how off-the-shelf telecommunication components can be incorporated into it for implementing an efficient quantum circuit.

In our scheme (See Fig.~\ref{Fig3}), the light source is composed of a semiconductor distributed feedback telecom laser ($\lambda = 1546$ nm) connected to an external fiber-pigtailed amplitude modulator (FMZ). Driven by a field programmable gate array electronic unit (FPGA1), we use the FMZ to externally modulate the laser to generate optical pulses 5 ns wide. Optical attenuators (ATT) are then used to create weak coherent states \cite{Gisin_2002}.

Following the attenuator, we use a commercial spatial demultiplexer/multiplexer unit (DEMUX) \cite{Watanabe_2012,Tottori_2012}, with insertion losses around $3.2 \%$. This device is composed of four independent single-mode fibers connected to a 4-core MCF. Each single-mode fiber is mapped to one of the cores of the MCF fiber. In our system, after the first DEMUX, only one of the MCF cores is illuminated, which is shown schematically in Fig.~\ref{Fig3}. This MCF fiber is then connected to a 4$\times$4 MBS as the starting point of the programmable 4-arm MZ interferometer. A second DEMUX unit (identical to the initial one but connected in reverse) is then used to separate the cores in individual single-mode fiber outputs, allowing access to each core. Each path contains two fiber-pigtailed LiNbO$_3$ phase modulators (PM) with 10 GHz bandwidth. This allows us to prepare and measure a more general class of path qudit states. Each PM has an internal polariser used to align the photon polarisation state such that in the interferometer there is no path-information available \cite{QErasure,DecQErasure}, which would compromise the visibility of the observed interference.  Fiber-based polarization controllers (not shown) are used in each path to maximize transmission through the PMs.  The first set of PMs is also controlled by the FPGA1 unit and are used for state preparation. The general form of the states that are prepared is
\begin{equation}
 |\chi\rangle= \frac{1}{2}(e^{i\phi_{0}^{A}} |0 \rangle+e^{i\phi_{1}^{A}} |1 \rangle+e^{i\phi_{2}^{A}} |2 \rangle+e^{i\phi_{3}^{A}} |3 \rangle),
\end{equation} where $\phi_k^{A}$ is the phase applied by the first modulator in mode $k$.

Finally, the state projection is done by another 4$\times$4 MBS, whose input is first converted from the four individual single-mode arms to a single 4-core fiber by a third DEMUX unit. Considering the 4$\times$4 MCF MBS matrix representation, and the action of the second set of PMs, one can show that the form of the measurement basis states at the end of the circuit are given by
\begin{eqnarray}
|\psi_{0}\rangle= \frac{1}{2}(e^{i\phi_{0}^{B}} |0 \rangle+e^{i\phi_{1}^{B}} |1 \rangle+e^{i\phi_{2}^{B}} |2 \rangle+e^{i\phi_{3}^{B}} |3 \rangle) , \nonumber \\
|\psi_{1}\rangle= \frac{1}{2}(e^{i\phi_{0}^{B}} |0 \rangle+  e^{i\phi_{1}^{B}} |1 \rangle- e^{i\phi_{2}^{B}} |2 \rangle- e^{i\phi_{3}^{B}} |3 \rangle), \nonumber \\
|\psi_{2}\rangle= \frac{1}{2}(e^{i\phi_{0}^{B}} |0 \rangle- e^{i\phi_{1}^{B}} |1 \rangle+e^{i\phi_{2}^{B}} |2 \rangle-e^{i\phi_{3}^{B}} |3 \rangle), \nonumber \\
|\psi_{3}\rangle= \frac{1}{2}(e^{i\phi_{0}^{B}} |0 \rangle-  e^{i\phi_{1}^{B}} |1 \rangle- e^{i\phi_{2}^{B}} |2 \rangle+ e^{i\phi_{3}^{B}} |3 \rangle),
\end{eqnarray} where $\phi_k^{B}$ is the phase applied by the second modulator in the core mode $k$. The second set of PMs is independently controlled by a second FPGA2 unit. In order to connect the second 4$\times$4 MBS to single-photon detectors ($D_i$) and conclude the measurement process, a fourth DEMUX unit is employed to split the 4-core fiber into four single-mode fibers. They are each connected to commercial InGaAs single-photon detection modules, working in gated mode and configured with 10$\%$ overall detection efficiency, and 5 ns gate width. The detectors' counts are simultaneously recorded by the FPGA2 unit. Through the control of the clock-synchronized FPGA units, one can program the generated path qudit states and measurements to be implemented by the circuit. Last, we note that while only three phase modulators are needed for each set, we opted to maintain the fourth one for easily matching the paths for future applications.

\begin{figure}[t]
\centering
\includegraphics[width=\linewidth]{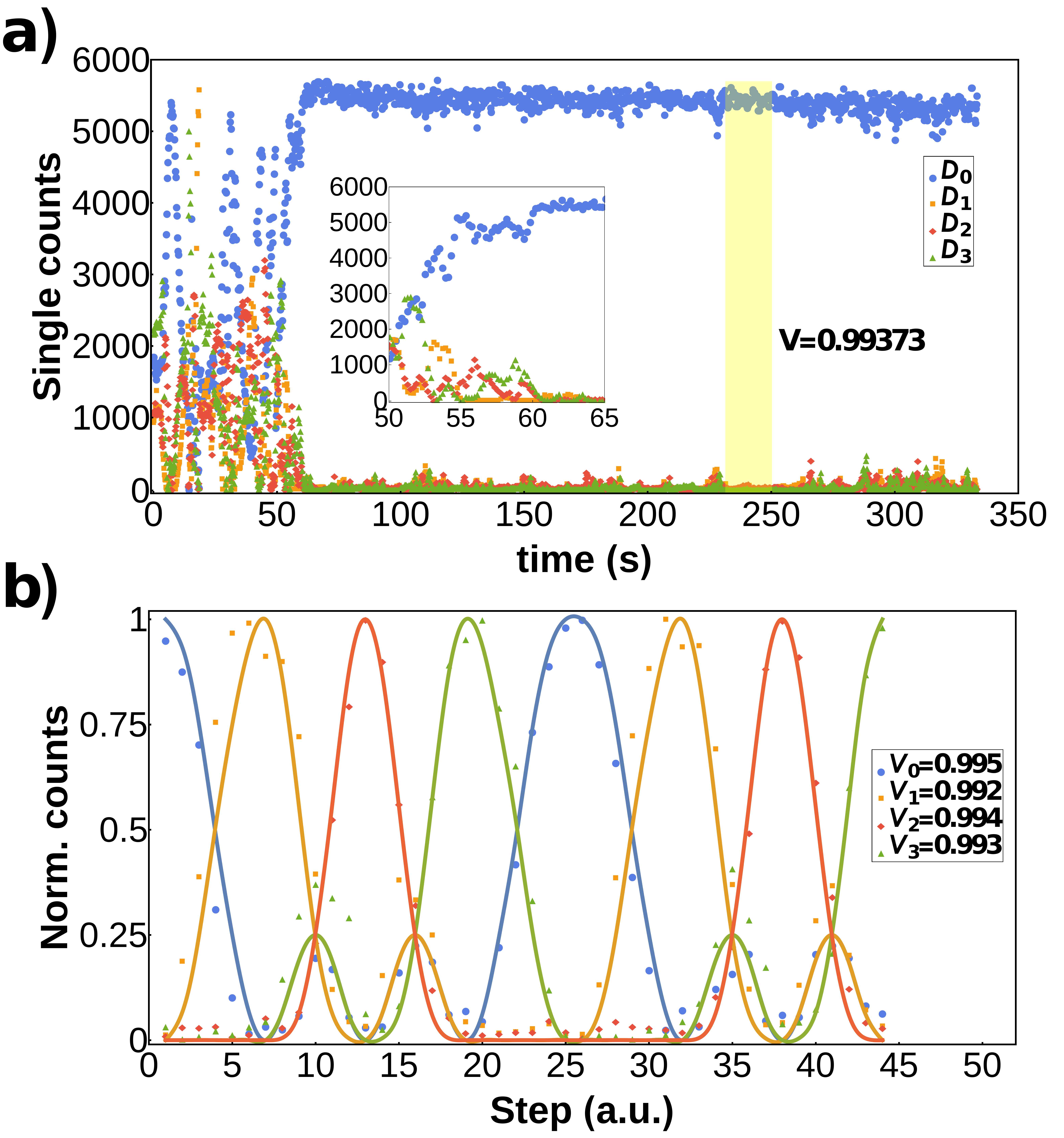}
\caption{Phase stabilization and interference fringes of the 4-arm programmable circuit. a) Active stabilization of the multi-arm interferometer (integration time 0.1s). Inset shows a zoom between 50 and 65 s showing the settling time of the control system after turning it on. b) Detection rate as a function of modulated phases $\phi^{A}_k$ (Integration time 1s).} \vspace{-0.2cm}\label{Fig4}
\end{figure}

The interferometer occupies a 30cm$\times$ 30cm area and is thermally insulated to minimize additional random phase drifts between the single-mode fibers. Nevertheless, long-term phase drifts are present and we implemented a control system to actively compensate them. The control is implemented by FPGA2 and it is based on a perturb and observe power point tracking method \cite{Nema}. More specifically, each applied phase $\phi_k^{B}$ can be decomposed as $\phi_k^{B} = \phi^{bias}_k+\phi^{mod}_k$, where the employed voltage driver is capable of supplying the sum of two independent voltages $V_{bias}$ and $V_{mod}$. $V_{bias}$ is a low-speed signal used to control $\phi^{bias}_k$, and this is intended to compensate a given phase drift $\phi^n_k$. $V_{mod}$ is the high-speed signal for modulating the desired phase $\phi^{mod}_k$. Since the total relative phase at the $k_{th}$ arm is $\phi_k^{B} = \phi^{bias}_k+\phi^{mod}_k + \phi^{n}_k$, the phase drift compensation algorithm running in FPGA2 will perturb the $k$th phase modulator to make $\phi^{bias}_k = - \phi^{n}_k$, such that the phase noise is eliminated. This is done by maximizing the number of photo counts at detector D$_0$, {which} corresponds to a situation where there is constructive interference. The algorithm does this sequentially to each phase modulator at the measurement stage. The multi-arm interferometer works with a repetition rate of 2 MHz and has an integration time of 0.1~s. When the system is initialized, the stabilization control typically takes around 15~s to align the interferometer as shown in {the experimental data in} Fig.~\ref{Fig4}a, {where the control system was activated} at $t=50$~s. When this point is achieved, the quantum circuit automatically prepares the desired states and performs the required measurements over experimental blocks of 0.1~s. The control system {monitors}  the phase stabilization of the interferometer {in real-time}, such that it stops the measurement procedure every 0.2~s to check the stabilization. The circuit can realign itself and run for several days continuously. To show the quality of the MCF based multi-arm interferometer, we gradually generate the quantum states associated to each outcome of the interferometer when all $\phi_k^{B}$s are set to zero, obtaining the traditional interference curves of Fig.~\ref{Fig4}b. The average visibility recorded is $0.992\pm 0.0015$, showing that path qudit states can be prepared and measured with high-fidelities in our scheme.

One last point is related with the overall detection efficiency of the circuit, which is a crucial parameter for many fundamental studies and applications in QI science. In our circuit, the transmission of the generated ququarts through the measurement stage was $(43 \pm 0.1)\%$, which is mainly limited by the second set of PMs that add an average 2.05 dB of insertion losses. Note, however, that this value represents a gain of up to 2 orders of magnitude compared with some aforemention HD experiments, where filtering techniques drastically reduce the transmission of the employed schemes (see \cite{Canas_2017}, for example). Considering that new commercially available superconducting detectors can reach more than 85$\%$ of detection efficiency, one can see that our system is capable of reaching at least $35\%$ of overall detection efficiency. Moreover, PMs with smaller insertion losses ($< 5\%$) based on poled fibres \cite{Walter1, Walter2} have recently been developed which could be incorporated to the system, and we estimate that an optimized circuit could reach 65$\%$ overall efficiency.

\section{Measurement-device-independent RNG: theory}

In the scenario of MDI RNG an end-user in need of random numbers possesses a characterised preparation device and an uncharacterised measurement device $\mathcal{M}$ \cite{Daniel2017}. This scenario is relevant nowadays as single-photon detectors are prone to side-channel attacks, which has motivated the development of similar approaches in quantum key distribution \cite{mdiQKD}. The preparation device is used to prepare quantum states, $\{\omega_x\}$, which are measured by the uncharacterized measuring device $\mathcal{M}$, leading to a classical outcome $a$. By repeating the process, one estimates the probabilities $p(a|\omega_x)$. Importantly, $\mathcal{M}$ could have been constructed by an eavesdropper (named Eve), who aims to predict the outcome $a$. Eve in principle can even be quantum-correlated with $\mathcal{M}$, by holding half of an entangled state $\rho^{AE}$, the other half of which is inside the device. $\mathcal{M}$ performs a measurement on the input state $\omega_x$ and a part of $\rho^{AE}$, while Eve makes a measurement on her part of $\rho^{AE}$ to guess the bit generated.

In \cite{Daniel2017} it was shown that the maximal probability $P_g(x^*)$ that Eve guesses correctly the outcomes $a$ for a given input $x^*$, compatible with $p(a|\omega_x)$, can be estimated by the solution of a semi-definite program \cite{SDP}. Finally the amount of randomness that is certified per round under the assumption that Eve carries out individual attacks is given by the min-entropy of $P_g$
\begin{equation}\label{MinEnt}
    H_{\min} (x^*) = -\log_2 P_g(x^*).
\end{equation}

A drawback of the approach proposed in \cite{Daniel2017} is that it relies on having exact knowledge of the probabilities $p(a|\omega_x)$. In any real experiment we only have access to a finite number of experimental rounds, which allows us to estimate the frequencies $\xi(a|\omega_x)$ that different measurement results are observed. To account for finite-statistics effects we adapt the semi-definite program described in \cite{Daniel2017} to make use of the Chernoff-Hoeffding tail inequality \cite{Hoeffding}. {This} inequality asserts that with high probability $p(a|\omega_x)$ is bounded by the observed frequencies $\xi(a|\omega_x)$ via
\begin{equation}\label{chernoff}
\xi(a|\omega_x) - t_x(\epsilon) \leq p(a|\omega_x) \leq \xi(a|\omega_x) + t_x(\epsilon),
\end{equation} where $ t_x(\epsilon) = \sqrt{\frac{\log(1/\epsilon)}{2n_x}}$ depends on a confidence parameter $\epsilon$ and the total number of measurement rounds $n_x$ in which the input was $\omega_x$. The confidence parameter corresponds to the probability that \eqref{chernoff} is not satisfied. In our analysis we choose $\epsilon = 10^{-9}$ (see supplementary material for details).

An implementation of the MDI RNG protocol with 4-dimensional quantum states involves the state preparation device that can randomly prepare five different states. Four of them,  $\{\ket{\omega_x}\}_{x=0}^3$, are orthogonal to each other, and the fifth $\ket{\omega_{4}}$, is mutually unbiased with respect to the first four, so that $|\langle\omega_x|\omega_4 \rangle|^2=1/4~\forall x=0,...,3$. The measuring device is set to measure in the basis spanned by $\{\ket{\omega_x}\}_{x=0}^3$, so that the measurement outputs are uniformly random whenever the state $\ket{\omega_{4}}$ is measured.  The min-entropy \eqref{MinEnt} for this ideal implementation gives $H_{min}(x=4)=2$, showing that $2$ bits of randomness per round can be generated.

\section{Measurement-device-independent RNG: implementation}

%************************************* FIG 4 ****************************************
\begin{figure}[t]\label{figura11Dic2018}
\centering
\includegraphics[width=\linewidth]{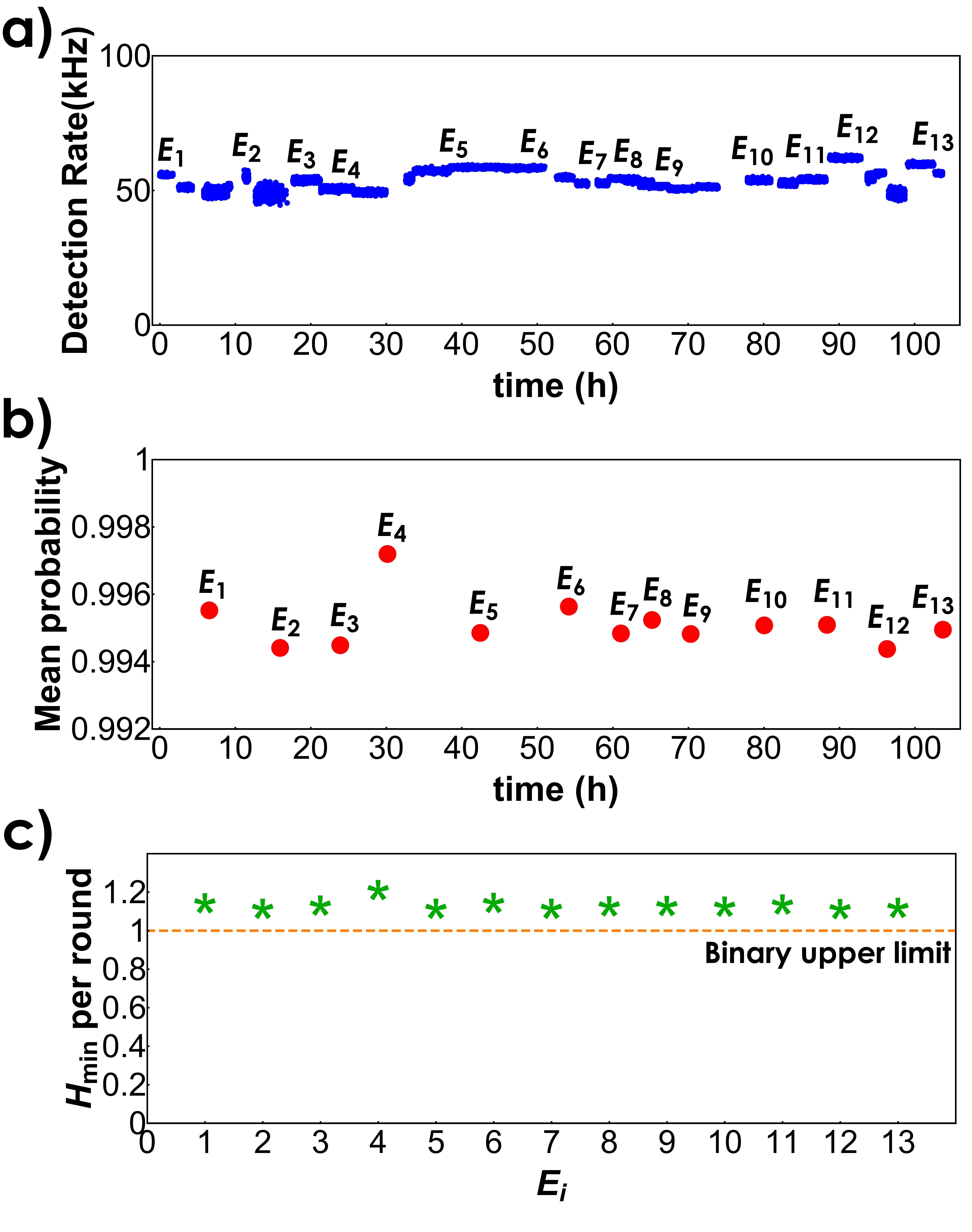}
\caption{Fragment of the data recorded over time. a) Single count detection rate considering only the selected samples (Please see text for details). $E_i$ with $i=\left\{ 1,2, .. ,13\right\}$ represent small zones, mostly between or with long realignment procedures.  b) Observed average success probability for each zone $E_i$. Error bars lie within the experimental point representation. c) Average obtained randomness per experimental round for each zone $E_i$. Error bars lie within the star symbols. The dashed line represents the theoretical upper bound allowed for binary RNG protocols. \label{Fig5}}
\end{figure}
%*************************************************************************************

As previously explained, our source consists of an attenuated pulsed laser that produces weak coherent states. The probability of emitting $j$ photons per pulse is characterised by the mean photon number, $\mu$, such that $p(j) = e^{-\mu}\mu^j/j!$. We consider states with average mean photon numbers of $\mu=0.2$ and $\mu=0.4$, while recording the single, double and triple coincidences counts between the four detectors $D_i$. Typically, for the experiment working with $\mu=0.4$, we observe $\sim 50000\pm 225$ single counts per second, $\sim 90\pm 9$ double-coincidences and only $1\pm 1$ triple-coincidence count. For $\mu=0.2$ we have not observed any triple coincidences. Thus, in our randomness analysis we consider a multi photon Hilbert space truncated up to two photons. Moreover, we adopt the fair sampling assumption and post-select on having at least one photon detected. Then, the set of input states has the following form:
\begin{align}
&\rho_x = p(1)\ket{\omega_x}\bra{\omega_x} + p(2)\ket{\phi_x^{(2)}}\bra{\phi_x^{(2)}},
\end{align}
where $p(1)+p(2)=1$, $\ket{\omega_{x=0}}=\ket{0}$,...,$\ket{\omega_{x=3}}=\ket{3}$ are the states corresponding to one photon travelling {in each mode (labelled by $x$)}, $\ket{\omega_4}=(\ket{0}-\ket{1}+\ket{2}+\ket{3})/2$ is the mutually unbiased state, and $\ket{\phi_x^{(2)}}$ refers to states where $2$ photons are generated in a single pulse. These states are given in the Supplementary Material. Notice that, in our experiment we observe ten measurement outcomes: four single clicks corresponding to photon detection at one of the four detectors $D_i$ ($i=0,...,3$), and six coincidence detections between detectors $D_i$ and $D_j$ with $i\neq j$. The statistics of all these events are taken into consideration in the randomness estimation.

The experiment operates at the repetition rate of 2 MHz. Over the course of one integration sample of 0.1s, $90\%$ of the rounds are randomly chosen by FPGA1 to send $\rho_{4}$. The other $10\%$ of samples are uniformly chosen between $\rho_{x=0}$,...,$\rho_{x=3}$. In this way, we prioritize the generation of random bits, while still having enough statistics to certify the amount of private randomness created. We continuously verify that the protocol is working properly through the average success probability of identifying the states $\omega_x$ (i.e., $\bar{p}= \frac{1}{4} \sum_{x=0}^{3} p(x|\rho_x$). If $\bar{p}>0.992$, then the random bit sequence is recorded. Otherwise, the control system starts a realignment procedure automatically. This threshold value has been chosen to maintain the system producing more than 1 bit of randomness per experimental round, the maximum that a RNG protocol based on dichotomic outcomes (and post-processing of it) would achieve.

Fig. \ref{Fig5} shows a fragment of the recorded data while the random number generator was operating with $\mu=0.4$. The points in Fig. \ref{Fig5}a represent the single photon detection rate in kHz. There are discontinuities that arise from the fact that only the results when $\bar{p} >0.992$ are displayed. $E_i$ with $i=\left\{ 1,2, .. ,13\right\}$ represent small zones, {between which the realignment procedure occurs}. The system is continuously realigning itself, but sometimes it does not quickly achieve a visibility higher than the given threshold. The experiment ran over a total of 103.7 hours. The corresponding average success probabilities per zone $E_i$ are shown in Fig. \ref{Fig5}b. The total average success probability is $\bar{p_t}=0.9946 \pm 0.0001$. From all the recorded data, the minimum entropy is estimated. The experimental $H_{min}^{exp}$ is bounded by $1.133<H_{min}^{exp}<1.232$, with its maximum value obtained at zone $E_4$ (See Fig. \ref{Fig5}c). The average value is $\bar{H}_{min}^{exp}=1.153 \pm 0.007$, which implies that the generator works with an average private random bit key rate of $\sim 57650\pm 350${bits}/s. With additional improvements in temporal width of the pulses, and faster clock rates of the detectors, it should be possible to increase this by at least two orders of magnitude. For the case with $\mu=0.2$ we obtain similar results. In this case, $H_{min}^{exp}$ is bounded by $1.134<H_{min}^{exp}<1.178$, with the average value given by $\bar{H}_{min}^{exp}=1.156 \pm 0.003$. Thus, we have demonstrated the robustness of the MDI RNG method while being implemented with weak coherent states. Importantly, these results show that the random number generator has been able to exploit the advantages provided by HD quantum systems, since it always produces a min-entropy greater than 1 bit per experimental round. We notice that a theoretical upper bound to the private random bit key rate is given by the min entropy of the most likely measurement outcome, which corresponds to an attack where an Eavesdropper always bets on this outcome. In our case, this corresponds to $H^{the}_{min} \approx 2.03$ for $\mu=0.4$ and $H^{the}_{min} \approx 2.02$ for $\mu=0.2$ (see supplementary material).

\section{Conclusion}

We have reported on the production and characterization of high-quality $N\times N$ multi-port beam-splitter devices built directly within a multi-core fiber. This is an important step towards the construction of universal photonic quantum information processing circuits based entirely on the new multi-core fiber platform, which will take advantage of the high phase stability provided by these fibers. We use a $4 \times 4$ device to experimentally show that a programmable quantum circuit for efficient 4-dimensional quantum information processing can be built using multi-core fiber based technology. Since it is constructed with commercially available components, it can be easily integrated with telecom fiber networks. To demonstrate the versatility and advantages of this circuit, we have demonstrated a measurement-device-independent quantum random number generator using 4-dimensional photonic states, which yielded a maximum of 1.23 private certified random bits generated per experimental round, surpassing the 1-bit limit of binary protocols. To achieve these results we employ a theoretical approach that allows for the evaluation of available private randomness using semi-definite programming and taking into account finite statistics of events. Furthermore, our programmable circuit operates at 2 MHz repetition rate (scalable to several GHz), generating about $6\times 10^4$ random bits/s. With scalability taken into account, our results compare favorably in terms of generation rate to other state-of-the-art quantum certified randomness generators, while providing better scalability to even higher dimensions. These results are critical in demonstrating a new robust and versatile high-dimensional quantum information processing platform for implementing universal programmable optical circuits.  In this regards, note that MCF BS technology has very recently been used to implement a quantum computational circuit based on a quantum $N$-switch \cite{Taddei_2020}.

\section*{Funding Information}
Fondo Nacional de Desarrollo Científico y Tecnológico (1200859, 1200266, 1190933, 3170596, 3170400); Millennium Institute for Research in Optics, MIRO. Ram\'on y Cajal fellowship, Spanish MINECO (QIBEQI FIS2016-80773-P and Severo Ochoa SEV-2015-0522). AXA Chair in Quantum Information Science, Generalitat de Catalunya (SGR875 and CERCA Programme). Fundaci\'{o} Privada Cellex. ERC CoG QITBOX. Royal Society University Research Fellowship (UHQT). COST project CA16218, NANOCOHYBRI. Brazilian grants CNPq 304196/2018-5,  FAPERJ E- 26/010.002997/2014 and E-26/202.7890/2017, and the INCT-Informa\c{c}\~ao Quântica. Ceniit Link\"{o}ping University and the Swedish Research Council (VR 2017-04470).

\section*{Disclosures}

The authors declare no conflicts of interest.

\section*{Supplemental Documents}

\subsection{Process Tomography}
	
In this section we explain the process tomography method introduced in \cite{Rahimi2013}, which we adopted for characterizing the $N \times N$ MCF MBS.  A unitary matrix is given by
	\begin{equation}
	U=\sum_{jk}u_{jk}e^{i\phi_{jk}}|j\rangle\langle k|.
	\end{equation}
	where $0\leq u_{jk}\leq 1$ and $0\leq \phi_{jk}< 2\pi$. The protocol is based on measuring the probabilities $I_{jk}$ of detecting the photons at the port represented by $|j\rangle$ when the core-mode state $|k\rangle$ was sent through the MBS. In this way, we determinate the split ratios
	\begin{equation}
	r(j|k) = \frac{I_{jk}}{\sum_k I_{jk}}.
	\end{equation}
	The parameters $u_{jk}$ are the square root of the split ratios,
	\begin{equation}
	u_{jk}=\sqrt{r(j|k)}  .
	\end{equation}
	On the other hand, the phases $\phi_{jk} $ are determined by sending states of the form
	\begin{equation}
	|\phi_j\rangle = \frac{1}{\sqrt{2}}\left(|1\rangle+e^{i\varphi}|j\rangle \right)
	\end{equation}
	through the MBS. The probability distribution $p(j|k)$ to detect photons at the output port $|k\rangle$ is given by
	\begin{equation}
	p(k|j)=\frac{1}{2}\left[u_{k1}^2+u_{kj}^2+2u_{k1}u_{kj}\cos\left(\varphi+\phi_{kj}-\phi_{k1}\right)  \right].
	\end{equation}
	Hence, since we know the coefficient $u_{jk}$ form the split ratios, by recording the probabilities $p(k|j)$ with respect to $\varphi$, one can obtain the relative phases $\phi_{kj}-\phi_{k1}$. However, note that only $N^2-2N+1$ of the phases $\phi_{jk}  $ are physically significant since $2N-1$ phases can be included into the basis vectors or externally controlled by phase modulators (PM) \cite{Rahimi2013,Bernstein1974,Peres1989}. Therefore, without loss of generality, we can consider that $\phi_{0k}=\phi_{j0}=0$, or equivalently, the matrix $U$ has real border. The procedure allows us to determine uniquely the phases $\phi_{jk}$, and one can obtain an experimentally estimated matrix $\tilde{U}$ of the MCB. Nonetheless, due to inherent experimental errors, this matrix is never unitary. In order to obtain the unitary matrix describing the $N \times N$ MCF MBS, one can optimize a cost function of the experimental data. For this purpose, we use the fidelity between two matrices \cite{Baldwin2014,Acin2001,Clements2016}, given by
	\begin{equation}
	F(A,B)=\frac{1}{N}\left|\mathrm{Tr}(A^\dagger B)\right|^2.
	\end{equation}
	This function is equivalent to the fidelity between the quantum states corresponding to $A$ and $B$ by the Choi-Jamiolkowski map. The unitary estimate of the MBS is obtained by solving the following optimization problem
	\begin{equation}
	\hat{U}=\arg\min_{V} [1-F(\tilde{U},V)],
	\end{equation}
	with the restriction that $ V $ is a real-border unitary matrix. This can be converted into an unconstrained optimization problem using the fact that, for an arbitrary complex matrix $Z$, we have that $Z(Z^\dagger Z)^{-1/2}$ is always a unitary matrix. Therefore, this optimization can be solved numerically by standard optimization methods.
	
\subsection{4x4 MCF MBS}
	The 4x4 MCF MBS is illustrated in Fig. 1 b) of the main text.
	The use of the protocol described before to characterize the 4x4 MCF MBS gives the following experimental estimate:
	\begin{equation}
	\tilde{U}_4 =
	\begin{bmatrix}
	0.5 & 0.5 &   0.5 &   0.5 \\
	0.5 &  -0.496 - 0.06i &  0.499 + 0.03i & -0.499 + 0.03i\\
	0.5 &   0.493 + 0.07i & -0.497 - 0.05i & -0.499 - 0.01i\\
	0.5 &  -0.5  & -0.496 + 0.06i &  0.499 - 0.03i
	\end{bmatrix},
	\end{equation}
	while its corresponding unitary one is
	\begin{equation}
	\hat{U}_4 =
	\begin{bmatrix}
	0.499 &0.501 &0.499 &0.499\\
	0.499 &-0.495-0.06i &0.498+0.03i &-0.499+0.03i\\
	0.501 &0.491+0.08i &-0.496-0.06i &-0.498-0.01i\\
	0.499 &-0.499-0.01i &-0.499+0.03i &0.499-0.01i
	\end{bmatrix}.
	\end{equation}
	We have that the fidelity between the experimental estimate and the unitary estimate is
	\begin{equation}
	%    F(\tilde{U}_4,\hat{U}_4)= 0.9995\pm0.0005.
	F(\tilde{U}_4,\hat{U}_4)= 0.999\pm0.001.
	\end{equation}
	Note that the unitary estimate is almost a symmetric unitary matrix, or equivalently, the absolute value of each coefficient of the matrix is approximately $1/2$. Comparing the unitary estimate with the symmetric unitary matrix
	\begin{equation}
	V_\phi = \frac{1}2\begin{bmatrix}
	1&1&1&1\\1&e^{\phi}&-1&-e^{\phi}\\1&-1&-1&1\\1&-e^{\phi}&1&-e^{\phi}
	\end{bmatrix}
	\end{equation}
	with $\phi=0$, we have the fidelity
	\begin{equation}
	%    F(\hat{U},V_{\phi=0})=0.9975\pm0.0015.
	F(\hat{U},V_{\phi=0})=0.995\pm0.003.
	\end{equation}
	
\subsection{7x7 MCF MBS}
	
	\begin{figure}
\centering
\includegraphics[width=0.6\linewidth]{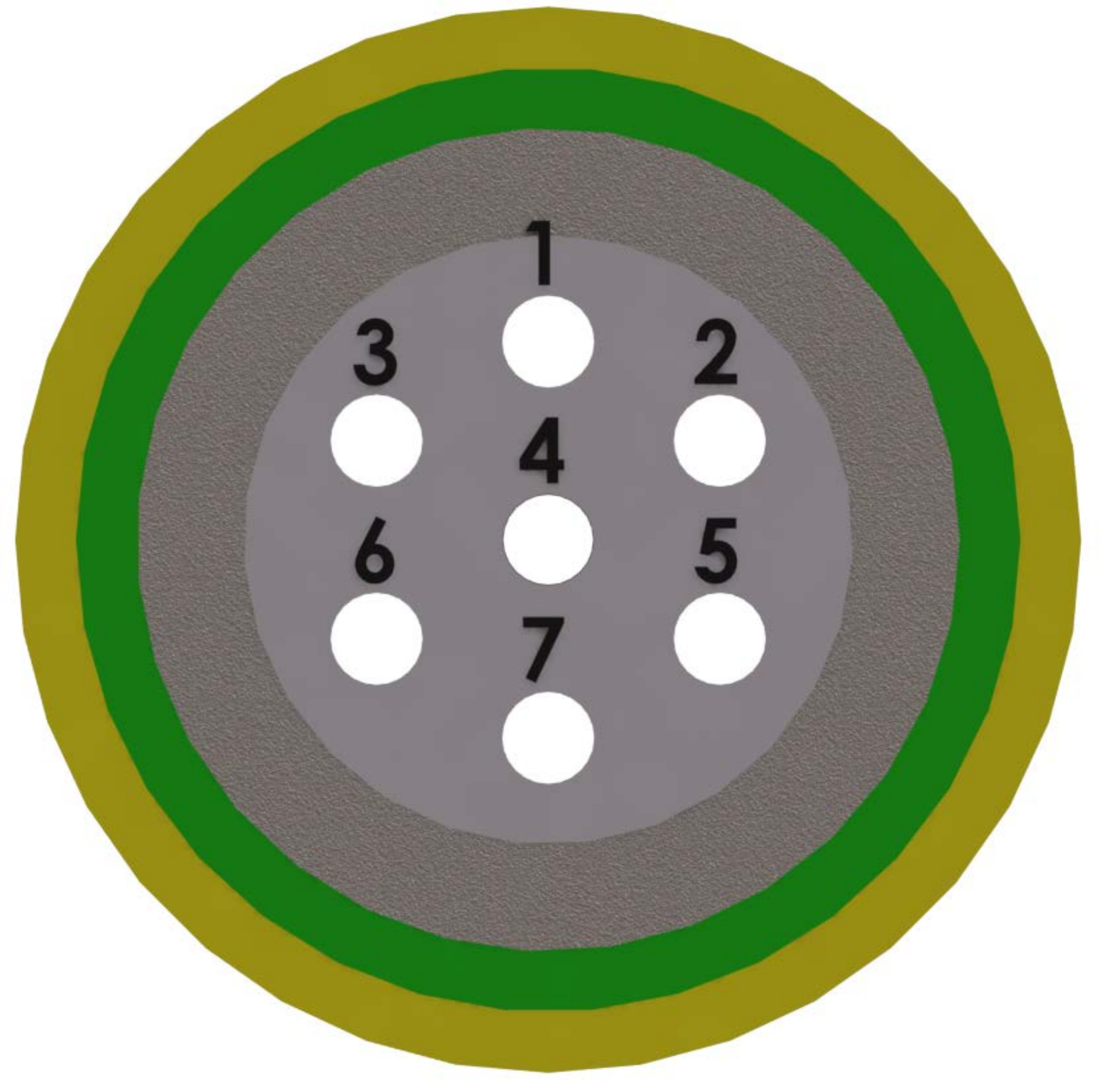}
\caption{Diagram showing geometry and labelling of paths in the $7 \times 7$ MCF MBS. \label{Fig7C}}
\end{figure}

	A diagram of the geometry of the $7 \times 7$ MCF MBS is shown in Fig. \ref{Fig7C}.  Using a procedure analogous to the $4 \times 4$ case, we obtain that experimental estimated matrix $\tilde{U}_7$ for the $7 \times 7$ MCF MBS:
\begin{widetext}	
\tiny	
\begin{equation}
	\tilde{U}_7 =
	\begin{bmatrix}
	0.5683 &   0.2049 &   0.3033 &   0.3795 &   0.4868 &   0.0949 &   0.4062 \\
	0.2214 &  -0.0031 + 0.1975i&  -0.5768 - 0.3410i&   0.3701 - 0.1266i&  -0.1344 + 0.4135i&   0.3159 - 0.2612i&  -0.1098 - 0.1671i\\
	0.3647 &  -0.6209 - 0.3041i&   0.1146 - 0.1043i&  -0.2181 - 0.2635i&  -0.0607 + 0.1064i&   0.0844 - 0.2467i&   0.0184 + 0.3920i\\
	0.3962 &   0.3325 + 0.0208i&  -0.1359 - 0.2890i&  -0.1493 + 0.3174i&  -0.0268 - 0.3410i&  -0.2991 - 0.1963i&  -0.3451 + 0.3673i\\
	0.3742 &  -0.1948 + 0.3179i&  -0.1422 + 0.1215i&  -0.0224 - 0.2837i&   0.0388 - 0.0591i&  -0.1498 + 0.7236i&  -0.3673 - 0.0641i\\
	0.1549 &   0.3464 + 0.1974i&   0.3318 - 0.1136i&  -0.0726 - 0.4156i&  -0.5162 + 0.2838i&  -0.2270 - 0.1024i&   0.1377 + 0.0065i\\
	0.4171 &  -0.0313 - 0.1790i&  -0.0587 + 0.4130i&  -0.3388 + 0.2987i&  -0.2960 - 0.0373i&   0.0918 - 0.0978i&   0.0971 - 0.4686i
	\end{bmatrix},\nonumber \label{tildeU7}
	\end{equation}
	\normalsize
	Following the procedure outlined above, the corresponding unitary matrix $\hat{U}_7$ is:
	\tiny
	\begin{equation}
	\hat{U}_7 =
	\begin{bmatrix}
	0.5639 &   0.2010 &   0.3019 &   0.3749 &   0.4918 &   0.0905 &  0.3998 \\
	0.2222 &  -0.0065 + 0.1874i&  -0.5700 - 0.3060i&   0.3558 - 0.0865i&  -0.1447 + 0.3632i&   0.2989 - 0.2884i&  -0.1033 - 0.1635i\\
	0.3487 &  -0.6271 - 0.3102i&   0.1178 - 0.0994i&  -0.2245 - 0.2686i&  -0.0469 + 0.1075i&   0.0629 - 0.2445i&  -0.0116 + 0.4061i\\
	0.3929 &   0.3320 + 0.0156i&  -0.1620 - 0.2950i&  -0.1267 + 0.3353i&  -0.0489 - 0.3414i&  -0.3319 - 0.1445i&  -0.3447 + 0.3530i\\
	0.3709 &  -0.1842 + 0.2868i&  -0.1199 + 0.1069i&  -0.0224 - 0.2699i&   0.0214 - 0.0533i&  -0.0144 + 0.7223i&  -0.3419 - 0.0698i\\
	0.1468 &   0.3709 + 0.2029i&   0.3572 - 0.0915i&  -0.0936 - 0.4318i&  -0.5262 + 0.3039i&  -0.2790 - 0.0553i&   0.1351 + 0.0108i\\
	0.4444 &  -0.0220 - 0.1704i&  -0.0651 + 0.4328i&  -0.3159 + 0.3254i&  -0.3157 - 0.0201i&   0.0839 - 0.1206i&   0.0989 - 0.4943i
	\end{bmatrix}.\nonumber \label{hatU7}
	\end{equation}
\normalsize
\end{widetext}
The fidelity between the experimental matrix and the unitary matrix is
	\begin{align}
	%F(\tilde U,\hat U)=0.9959\pm 0.0038.
	F(\tilde U,\hat U)=0.992\pm 0.008.
	\end{align}
	Note that this matrix is not symmetric, that is, its coefficients have absolute value different than $1/\sqrt{7}$. This is a consequence of the geometry of the cores in the fiber.
	
\subsection{Matrix Error Analysis}

We perform Monte Carlo simulations to quantify the error of the estimated matrices. We employ the Gaussian distribution $N(\mu,\sigma)$ for this task, where $\mu$ is the mean and $\sigma$ is the standard deviation. Considering the error as 3 times the standard deviation of the Gaussian distribution, approximately $99.7\%$ of the realizations are inside of the interval $\mu\pm 3\sigma$. Experimentally, we measure the intensities $I_{jk}\pm\Delta I_{jk}$ and the phases $\phi_{jk}\pm\Delta\phi_{jk}$, with $\Delta I_{jk}$ and $\Delta\phi_{jk}$ being their respective experimental errors.  Thereby, the simulated split ratios and phases are given by
	\begin{equation}
	\bar{I}_{jk}\sim N(I_{jk},\Delta I_{jk}/3) ,\qquad    \bar{\phi}_{jk}\sim N(\phi_{jk},\Delta{\phi_{jk}}/3),
	\end{equation}
	respectively. We generate a sample of $10^5$ MBS matrices $\tilde{U}$ and $\hat{U}$ independently, and with them we calculate the average fidelity and their respectively errors, which were consider as 3 times the standard deviations.

\subsection{MDI RNG protocol details}

In a MDI RNG scenario, an end-user possesses a characterised preparation device $\mathcal{P}$ used to prepare a set of quantum states $\{\omega_x\}$, which are measured by the uncharacterized measuring device $\mathcal{M}$, leading to a classical outcome $a$. It is assumed that an eavesdropper, Eve, can be quantum-correlated with $\mathcal{M}$, by holding half of an entangled state $\rho^{AE}$, the other half of which is inside the device. $\mathcal{M}$ performs a measurement (which can be known by Eve) on the input state $\omega_x$ and a part of $\rho^{AE}$, while Eve uses a positive operator valued measure (POVM) $N_e^E$ to measure her part of $\rho^{AE}$.

After many uses of the device the user estimates the probabilities $p(a|\omega_x)$.
In \cite{Daniel2017} it was shown that the maximal guessing probability of Eve, for a given input $x^*$, compatible with $p(a|\omega_x)$, is bounded by the solution of the following semi-definite program (SDP) \cite{SDP}
\begin{align}\label{sdp}
P_g(x^*) = \max&\!\!\!\!\!\quad \mathrm{tr} \sum_{a} N_{a,e=a} \omega_{x^*} \\ \nonumber
\text{s.t.}&\!\! \quad p(a|\omega_x) = \mathrm{tr}\sum_e N_{ae}\omega_x \quad \forall a,x \\ \nonumber
&\quad \sum_a N_{ae} = q(e) \mathbb{I} \quad \forall e \\ \nonumber
&\quad \sum_e q(e) = 1,
\end{align}
where the maximisation is over the POVM $\mathbb{N} = \{N_{ae}\}_{ae}$ and probability distribution $\mathbf{q} = \{q(e)\}_e$, and the second constraint encodes no-signalling between the measuring device and Eve (see \cite{Daniel2017} for details). The amount of randomness that it certified is given by the min-entropy,
\begin{equation}
    H_{\min} (x^*) = -\log P_g(x^*),
\end{equation}
assuming that Eve carries out individual attacks (i.e. does not share entanglement between rounds).

In order to account for effects due to finite statistics, we use the Chernoff-Hoeffding tail inequality \cite{Hoeffding}. It asserts that with high probability
\begin{equation}\label{chernoff}
\xi(a|\omega_x) - t_x(\epsilon) \leq p(a|\omega_x) \leq \xi(a|\omega_x) + t_x(\epsilon),
\end{equation}
where $\xi(a|\omega_x)$ are the frequencies observed in the experiment, and $ t_x(\epsilon) = \sqrt{\frac{\log(1/\epsilon)}{2n_x}}$ depends on a confidence parameter $\epsilon$ and the total number of measurement rounds $n_x$ in which the input was $\omega_x$. The confidence parameter corresponds to the probability that \eqref{chernoff} is not satisfied. A typical choice is to take $\epsilon = 10^{-9}$. Using this, \eqref{sdp} can be strengthened, so that it depends only upon the observed frequencies, namely
\begin{align}\label{sdpfs}
P_g(x^*) = &\max\!\!\!\!\!\quad \mathrm{tr} \sum_{a} N_{aa} \omega_{x^*} \\
\text{s.t.}& \!\!\!\! \quad  \xi(a|\omega_x)\! - \!t_x(\epsilon) \leq \mathrm{tr}\sum_e N_{ae}\omega_x \leq \xi(a|\omega_x)\! +\! t_x(\epsilon) \!\!\!\!\! \quad \forall a,x \nonumber \\
&\quad \sum_a N_{ae} = q(e) \mathbb{I} \quad \forall e \nonumber \\
&\quad \sum_e q(e) = 1.  \nonumber
\end{align}

\subsection{Experimental states}
In this section we give the states that are prepared in the experiment when two photons are emitted at the source. In the main text these states are labelled $\ket{\phi_x^{(2)}}$. In the main text, the notation used it that $\ket{x}$ referred to one photon in mode $x$. Here, since we want to have multiple photons in a given mode, we will use the notation $\ket{1000}$ to denote one photon in mode zero, $\ket{0200}$ to denote 2 photons in mode 1, etc. With this notation, the first four states have two photons in each mode, namely
\begin{widetext} \begin{align}
\ket{\phi_0^{(2)}} &= \ket{2000},& \ket{\phi_1^{(2)}} &= \ket{0200},& \ket{\phi_2^{(2)}} &= \ket{0020},& \ket{\phi_3^{(2)}} &= \ket{0002},
\end{align} 
while for the final state
\begin{multline}
\ket{\phi_4^{(2)}} = \frac{1}{\sqrt{28}}\Big(\ket{2000} + \ket{2000} + \ket{2000} + \ket{2000} - 2\left(\ket{0011} - \ket{0101} + \ket{0110} - \ket{1001} + \ket{1010} - \ket{1100}\right)\Big).
\end{multline} \end{widetext}

\end{document}